\documentstyle[amssymb,floats,aps,prb,epsf]{revtex}
\epsfclipon 
\begin{document}
\twocolumn[\hsize\textwidth\columnwidth\hsize\csname
@twocolumnfalse\endcsname 

\draft

\title{Coexistence of the electron Cooper pair and antiferromagnetic
short-range correlation in copper oxide materials}
\author{Shiping Feng}
\address{CCAST (World Laboratory) P. O. Box 8730, Beijing 100080,
China \\
Department of Physics, Beijing Normal University, Beijing
100875, China \\
Institute of Theoretical Physics, Academia Sinica, Beijing 100080,
China}

\maketitle

\begin{abstract}
Within the fermion-spin theory, the physical properties of the electron
pairing state in the copper oxide materials are discussed. According to
the common form of the electron Cooper pair, it is shown that there is
a coexistence of the electron Cooper pair and magnetic short-range
correlation, and hence the antiferromagnetic short-range correlation can
persist into the superconducting state. Moreover, the mean-field results
indicate that the electron pairing state originating from the pure
magnetic interaction in the two-dimensional $t$-$J$ model is the local
state, and then does not reveal the true superconducting ground-state.
\end{abstract}

]
\bigskip

\narrowtext

After over ten years of intense experimental studies of the copper oxide
superconductors, a significant body of reliable and reproducible data has
been accumulated by using many probes, which show that the properties of
the copper oxide superconductors can be explained in terms of the electron
pairing theory \cite{n1,n2}. The experimental evidence includes the factor
of $2e$ occurring in the flux quantum and in the Josephson effect, as well
as the electrodynamic and thermodynamic properties of the copper oxide
superconductors \cite{n3}. The experimental results from the angle resolved
photoemission spectroscopy imply that in the real space the gap function
and pairing force have a range of one lattice spacing \cite{n4,n41}. This
is much different from the superconductivity in the conventional metals,
where requires pairing with the long-range phase coherence \cite{n5}.
Moreover, the unusual normal state properties of the copper oxide
superconductors are also markedly different from those of the conventional
superconductors and are usually viewed as manifestations of strong
electron-electron correlations \cite{n6,n7}, which cause the
antiferromagnetic (AF) long-range-order (AFLRO) state in the undoped
copper oxide materials. Although this AFLRO disappears for the hole doping
concentration exceeds some critical value, the short-range AF correlation
still persist even into the superconducting state in the underdoped and
optimal doped regimes, which show that the short-range AF correlation may
play a role in both the unconventional normal state properties and the
mechanism of the superconductivity of the copper oxide materials
\cite{n6,n7}. Therefore there is the microscopic difference between the
conventional superconductors and copper oxide superconductors, namely
they have a different origin. It is believed that the correct theory for
describing the anomalous properties of the copper oxide materials should
involve the charge-spin separation in some form \cite{n8}.

Many researchers \cite{n9} have argued successfully that the $t$-$J$
model provides a consistent description of the physical properties of the
copper oxide materials. On the other hand, there is a lot of evidence
from the experiments and numerical simulations in favour of the $t$-$J$
model as the basic underlying microscopic model \cite{n6,n7,n91}. In
order to account for the real experiments based on the $t$-$J$ model, it
is crucial to impose the electron single occupancy on-site local
constraint \cite{n11}. To satisfy this local constraint in analytical
calculations, the fermion-spin theory has been proposed \cite{n12} to
study the $t$-$J$ model. In this approach, the physical electron is
decomposed into a spinless fermion and a hard-core boson, and then
naturally incorporates the physics of the charge-spin separation.
Within this approach, it has been shown \cite{n13}
that AFLRO in copper oxide materials vanishes around doping $\delta =5\%$.
The mean-field theory in the underdoped and optimal doped regimes without
AFLRO has been developed \cite{n14} to study the photoemission spectroscopy
and electron dispersion. Moreover, the charge
dynamics \cite{n15} and spin dynamics \cite{n16} in the normal state
of the copper oxide materials have been discussed by considering
fluctuations around this mean-field approximation (MFA), and the results
are in qualitative agreement with the experiments and numerical
simulations. In this paper, we apply this successful approach to discuss
the physical property of the electron pairing state in the copper oxide
materials. According to the common form of the electron Cooper pair, we
show that there is a coexistence of the electron Cooper pair and AF
short-range correlation, and hence the AF short-range magnetic
fluctuation can persist into the superconducting state. Moreover, it is
shown within the mean-field level that the singlet pair of electrons
originating from the pure magnetic interaction in the two-dimensional
(2D) $t$-$J$ model is local, and then does not reveal the true
superconducting ground-state.

According to the fermion-spin formulism \cite{n12}, the constrained
electron operators in the $t$-$J$ model can be decomposed as,
$C_{i\uparrow}=h^{\dagger}_{i}S^{-}_{i}$,
$C_{i\downarrow}=h^{\dagger}_{i}S^{+}_{i}$
with the spinless fermion operator $h_{i}$ keeping track of the charge
(holon), while the pseudospin operator $S_{i}$ keeping track of the spin
(spinon). In this representation, the $t$-$J$ model can be expressed
\cite{n12} as,
\begin{eqnarray}
H&=&t\sum_{\langle ij\rangle}h^{\dagger}_{j}h_{i}(S^{+}_{i}
S^{-}_{j}+S^{-}_{i}S^{+}_{j})-\mu \sum_{i}h^{\dagger}_{i}h_{i}
\nonumber \\
&+& J\sum_{\langle ij\rangle}(h_{i}h^{\dagger}_{i})({\bf S}_{i}
\cdot {\bf S}_{j}-{1\over 4})(h_{j}h^{\dagger}_{j}).~~~~~
\end{eqnarray}
We treat this Hamiltonian within the MFA by introducing the following
order parameters: $\chi =\langle S^{+}_{i}S^{-}_{i+\eta}\rangle$,
$\chi_{z} =\langle S^{z}_{i}S^{z}_{i+\eta}\rangle$,
$\phi_{\eta} =\langle h^{\dagger}_{i}h_{i+\eta}\rangle$, and
$\Delta^{(h)}_{\eta} =\langle h_{i}h_{i+\eta}\rangle$ with
$\eta=\pm \hat{x},\pm \hat{y}$. $\chi$ and $\chi_{z}$ describe
the spinon pair correlation, $\phi_{\eta}$ is the holon particle-hole
parameter, while $\Delta^{(h)}_{\eta}$ represents the holon pair
correlation. In this case, the $t$-$J$ Hamiltonian (1) can be decoupled
as $H=H_{t}+H_{J}-8Nt\chi\phi-4NJ_{eff}(\chi+\chi_{z})-4NV_{eff}(|
\phi_{\eta}|^{2}-|\Delta_{\eta}|^{2})$ with
\begin{eqnarray}
H_{t}&=&2\sum_{i,\eta}(\chi t+V_{eff}\phi_{\eta})
h^{\dagger}_{i+\eta}h_{i}-\mu\sum_{i}h^{\dagger}_{i}h_{i}
\nonumber \\
&-&V_{eff}\sum_{i,\eta}(\Delta^{(h)}_{\eta}h^{\dagger}_{i}
h^{\dagger}_{i+\eta}+\Delta^{(h)*}_{\eta}h_{i+\eta}h_{i}), \\
H_{J}&=&{1\over 2}J_{eff}\epsilon \sum_{i,\eta}(S^{+}_{i}
S^{-}_{i+\eta}+S^{-}_{i}S^{+}_{i+\eta}) \nonumber \\
&+&J_{eff} \sum_{i,\eta}S^{z}_{i}S^{z}_{i+\eta},
\end{eqnarray}
where $N$ is the number of sites, the doping dependent magnetic
exchange energy
$J_{eff}=J[(1-\delta)^{2}-\phi^{2}_{\eta}+|\Delta^{(h)}_{\eta}|^{2}]$,
the holon's effective attractive interaction
$V_{eff}=-J(\chi+\chi_{z}-1/4)$, and $\epsilon=1+2t\phi_{\eta}/J_{eff}$.
Since the holon order parameters $\phi_{\eta}$ and $\Delta^{(h)}_{\eta}$
are decoupled from the term
$V_{eff}\sum_{i,\eta}h_{i}h^{\dagger}_{i}h_{i+\eta}h^{\dagger}_{i+\eta}$
in the MFA, then there are many ways to choose the gauge for these order
parameters $\phi_{\eta}$ and $\Delta^{(h)}_{\eta}$. However, in the
copper oxide superconductors, some experiments seem consistent with
an s-wave pairing \cite{n2}, while other measurements gave the evidence
in favor of the d-wave pairing \cite{n1}. Therefore in the following
discussions, we only consider the cases of the s-wave pairing
$\phi_{y}=\phi_{x}=\phi$,
$\Delta^{(h)}_{y}=\Delta^{(h)}_{x}=\Delta^{(s)}_{h}$, and the d-wave
pairing $\phi_{y}=\phi_{x}=\phi$,
$\Delta^{(h)}_{y}=-\Delta^{(h)}_{x}=\Delta^{(d)}_{h}$, respectively.
In this case, the spinon mean-field Green's function $D(i-j,t-t^{\prime}
)=\langle\langle S^{+}_{i}(t);S^{-}_{j}(t^{\prime})\rangle\rangle$
of the Hamiltonian (3) have been discussed in detail in Ref. \cite{n14}
based on the Tyablikov \cite{n18} and Kondo and Yamaji \cite{n19} scheme.
For the convenience of the further discussions in this paper, this spinon
Green's function is rewritten here,
\begin{eqnarray}
D({\bf k},\omega)&=&{\Lambda [(2\epsilon\chi_{z}+\chi)\gamma_{k}-
(\epsilon\chi+2\chi_{z})]\over 2\omega (k)}\nonumber \\
&\times& \left ({1\over \omega -\omega (k)}-{1\over \omega +
\omega (k)}\right ) ,
\end{eqnarray}
where $\Lambda =8J_{eff}$,
$\gamma_{{\bf k}}=({\rm cos}k_{x}+{\rm cos}k_{y})/2$,
and the spinon spectrum $\omega(k)$ has been given in Ref.
\cite{n14}. While the holon mean-field Hamiltonian (2)
can be diagonalized by the Bogoliubov transformation in the momentum
space, and then the holon mean-field Green's function
$g({\bf k}, t-t^{\prime})=\langle\langle h_{k}(t);h^{\dagger}_{k}
(t^{\prime})\rangle\rangle$ and holon mean-field anomalous Green's
function $\Im_{h}({\bf k},t-t^{\prime})=-\langle\langle h_{-k}(t);h_{k}
(t^{\prime})\rangle\rangle$ are obtained as
\begin{eqnarray}
g({\bf k},\omega)&=&{1\over 2}\left (1+{\xi_{k}\over E_{k}}\right )
{1\over \omega-E_{k}}\nonumber \\
&+&{1\over 2}\left (1-{\xi_{k}\over E_{k}}\right )
{1\over \omega+E_{k}}, \\
\Im_{h}({\bf k},\omega)&=&-{\Delta^{(a)}_{h}(k)\over 2E_{k}}
\left ({1\over \omega-E_{k}}-{1\over \omega+E_{k}}\right ),
\end{eqnarray}
respectively, where the holon spectrum
$\xi_{k}=8(\chi t+\phi V_{eff})\gamma_{k}-\mu$,
and the quasi-particle excitation energy
$E_{k}=\sqrt{\xi^{2}_{k}+[\Delta^{(a)}_{h}(k)]^{2}}$.
The gap function
$\Delta^{(a)}_{h}(k)=8V_{eff}\Delta^{(a)}_{h}\gamma^{(a)}_{k}$ with
$\Delta^{(a)}_{h}=\Delta^{(s)}_{h}$, $\gamma^{(a)}_{k}=\gamma^{(s)}_{k}
=\gamma_{k}$ for s-wave pairing, and $\Delta^{(a)}_{h}=
\Delta^{(d)}_{h}$, $\gamma^{(a)}_{k}=\gamma^{(d)}_{k}=
({\rm cos}k_{x}-{\rm cos}k_{y})/2$ for d-wave pairing, respectively,
and satisfy the following equation,
\begin{eqnarray}
\Delta^{(a)}_{h} &=&{1\over 2N}\sum_{k}\gamma_{k}{\Delta^{(a)}_{h}(k)
\over E_{k}}{\rm th}[{1\over 2}\beta E_{k}].
\end{eqnarray}
This gap equation must be solved simultaneously with other self-consistent
equations \cite{n14}, therefore the holon pair transition temperature
$T^{(a)}_{(h)c}$ is determined by the condition
$\Delta^{(a)}_{h}(T^{(a)}_{(h)c})=0$.

Before discussing the properties of the electron pairing state, we first
consider the properties of the holon pairs. Although it is not observable
from experiments, its feature will
affect the properties of the physical electrons because of the strong
interaction between holons and spinons within the framework of the
charge-spin separation. Fig. 1 shows the results of the holon gap
parameter in the s-wave symmetry $\Delta^{(s)}_{h}$ (solid line) and
d-wave symmetry $\Delta^{(d)}_{h}$ (dashed line) as a function of
dopings at the temperature $T=0$ for (a) $t/J=2$ and (b) $t/J=2.5$,
where holons favour the s-wave pairing at low dopings and d-wave pairing
at high dopings. $\Delta^{(s)}_{h}$ ($\Delta^{(d)}_{h}$) is decreased
(increased) with increasing dopings, and the value of these gap
parameters is rather sensitive to the parameter $t/J$ and always
decreased with increasing $t/J$. It is surprised that the ranges of
the holon s-wave pairing state and holon d-wave pairing state are
also strong dependent on $t/J$, the boundary of the holon s-wave
pairing state is biased towards the lower dopings, while the boundary
of the holon d-wave pairing state is moved to the higher dopings with
increasing $t/J$, and we therefore find that in the underdoped and
optimal doped regimes ($0.05<\delta >0.25$) there are no in existence
of the holon d-wave pairing state for $t/J>2.3$, and the holon s-wave
pairing state for $t/J>3.7$. This indicates that the holon pairing
state induced by the pure magnetic interaction in the 2D $t$-$J$ model
is the local state,
since the $t$-$J$ model is characterized by a competition between the
kinetic energy ($t$) and magnetic energy ($J$), increasing the value
of the parameter $t/J$ means to increase the kinetic energy and tends
to destroy any local state. In the overdoped case ($\delta>0.25$), the
copper oxide materials become the better metals, and are described by
the Fermi-liquid theory, then the fermion-spin theory based on the
charge-spin separation breaks down. Therefore the above discussions
about the holon pairing state are only valid for the underdoped and
optimal doped regimes.

The holon pairing state originating from the magnetic interaction will
also lead to form the electron pairing state. The order parameter for
the electron pair is expressed as,
\begin{eqnarray}
\Delta_{\eta}=\langle C^{\dagger}_{i\uparrow}C^{\dagger}_{i+\eta
\downarrow}-C^{\dagger}_{i\downarrow}C^{\dagger}_{i+\eta\uparrow}
\rangle,
\end{eqnarray}
which describes the electron Cooper pair in a range of one lattice
spacing. In our present theoretical framework, the symmetry of the
electron gap parameter $\Delta_{\eta}$ is determined
by the symmetry of the holon gap parameter $\Delta^{(h)}_{\eta}$. For
discussing the physical properties of the electron pairing state, we
need to calculate the electron anomalous Green's function
$\Im^{\dagger}({\bf k},t-t^{\prime})=-\langle\langle C^{\dagger}_{k
\uparrow}(t);C^{\dagger}_{-k\downarrow}(t^{\prime})\rangle\rangle$,
which is a convolution of the spinon Green's function
$D({\bf k},t-t^{\prime})$ and anomalous holon Green's function
$\Im_{h}({\bf k},t-t^{\prime})$ in the framework of the fermion-spin
theory, {\it i.e.}, $\Im^{\dagger}({\bf k},t-t^{\prime})=(1/N)
\sum_{{\bf p}}D({\bf p},t-t^{\prime})\Im_{h}({\bf p-k},t^{\prime}-t)$,
and can be obtained at the mean-field level as,
\begin{eqnarray}
\Im^{\dagger}({\bf k},\omega )&=&-{1\over N}\sum_{p}{\Delta^{(a)}_{h}
(p-k)\over 2E_{p-k}}\nonumber \\
&\times& {\Lambda [(2\epsilon\chi_{z}+\chi)\gamma_{p}
-(\epsilon\chi+2\chi_{z})]\over 2\omega (p)}\nonumber \\
&\times&\left ( {F_{1}({\bf k},{\bf p})\over \omega -\omega (p)+
E_{p-k}}\right. \nonumber \\
&+&\left. {F_{2}({\bf k},{\bf p})\over \omega +\omega (p)+E_{p-k}}
\right .\nonumber\\
&-&\left . {F_{1}({\bf k},{\bf p})\over\omega +\omega (p)-E_{p-k}}
\right .\nonumber \\
&-& \left . {F_{2}({\bf k},{\bf p})\over\omega -\omega (p)-E_{p-k}}
\right),
\end{eqnarray}
where $F_{1}({\bf k},{\bf p})=n_{B}(\omega_{p})+n_{F}(E_{p-k})$,
$F_{2}({\bf k},{\bf p})=1+n_{B}(\omega_{p})-n_{F}(E_{p-k})$, with
$n_{B}(\omega_{p})$ and $n_{F}(E_{p-k})$ are the spinon and
holon distribution functions, respectively. In this case, the
electron gap equation is obtained according to the above anomalous
electron Green's function as,
\begin{eqnarray}
\Delta^{(a)}&=&-{2\over N}\sum_{k}\gamma^{(a)}_{k}{1\over N}\sum_{p}
\left ({\Delta^{(a)}_{h}(p-k)\over 2E_{p-k}}{\rm th}[{1\over 2}
\beta E_{p-k}]\right )\nonumber\\
&\times& {\Lambda [(2\epsilon\chi_{z}+\chi)\gamma_{p}-(\epsilon\chi
+2\chi_{z})]\over 2\omega (p)} \nonumber \\
&\times& {\rm coth}[{1\over 2}\beta\omega(p)],
\end{eqnarray}
which shows that the electron gap parameter $\Delta^{(a)}$ is strong
dependent on the holon gap parameter $\Delta^{(a)}_{h}$. It has been
shown \cite{n16} that the magnetic fluctuation is dominated by the
scattering of spinons, while in the present case, this magnetic
fluctuation has been incorporated into the electron anomalous Green's
function (and hence the electron Cooper pair) in terms of the spinon
Green's function $D({\bf k},\omega)$. Since the form of the electron
Cooper pair (8) is common, and the anomalous electron Green's function
always is the convolution of the spinon Green's function and anomalous
holon Green's function in the framework of the fermion-spin theory,
therefore there is the coexistence of the electron Cooper pair and
magnetic fluctuation, and hence the AF short-range correlation can
persist into the superconductivity, which is consistent with the
experiments \cite{n6}. Moreover, we find that although there is the
coexistence of the electron Cooper pair and magnetic fluctuation,
the value of the electron gap parameter still suppressed by this
magnetic fluctuation. Fig. 2 shows (a) the value of the electron gap
parameter $\Delta^{(a)}$ at the zero temperature and (b) electron pair
transition temperature $T^{(a)}_{c}$ as a function of doping $\delta$
for the s-wave symmetry (solid line) and d-wave symmetry (dashed line)
in $t/J=2.0$. In comparison with Fig. 1 (a), the value of $\Delta^{(a)}$
has been decreased to about half of $\Delta^{(a)}_{h}$, while the
electron pair transition temperature $T^{(s)}_{c}$ ($T^{(d)}_{c}$) is
almost identical to the holon pair transition temperature
$T^{(s)}_{(h)c}$ ($T^{(d)}_{(h)c}$), since according to Eq. (10), the
electron pair transition temperature obtained from the condition
$\Delta^{(a)}(T^{(a)}_{c})=0$ is essential same with these from
$\Delta^{(a)}_{h}(T^{(a)}_{(h)c})=0$. It has been shown from the
experiments \cite{n191} that as the doping concentration $\delta$
is reduced from the optimal doping, the superconducting gap is
constant or may be slightly increasing. This reflects that the true
superconducting gap is insensitive to the parameter $t/J$ in the $t$-$J$
model. However, in the present case the electron pairing state is
dominated by the holon pairs, and as our above discussions for the
properties of the holon pairing state, the present electron pairing
state originating from the pure magnetic interaction in the 2D $t$-$J$
model also is the local state, and then does not reveal the true
superconducting ground-state, which is consistent with the numerical
result obtained by Shih {\it et al.} \cite{n192}, they show that the
pure 2D $t$-$J$ model does not have long-range d-wave superconducting
correlation in the interesting parameter range of $t/J\geq 2$.

\begin{figure}[prb]
\epsfxsize=3.0in\centerline{\epsffile{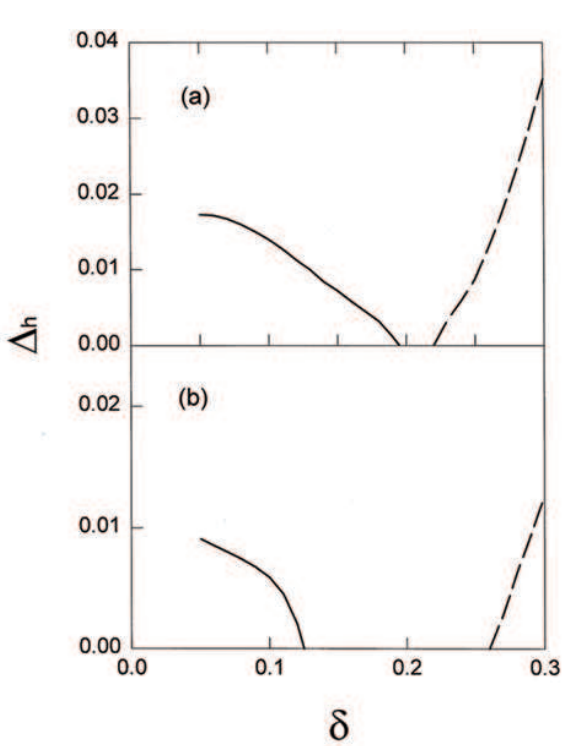}}
\caption{The holon gap parameter in the s-wave symmetry
$\Delta^{(s)}_{h}$ (solid line) and d-wave symmetry
$\Delta^{(d)}_{h}$ (dashed line) as a function of doping $\delta$
at the zero temperature for (a) $t/J=2$ and (b) $t/J=2.5$.}
\end{figure}

\begin{figure}[prb]
\epsfxsize=3.0in\centerline{\epsffile{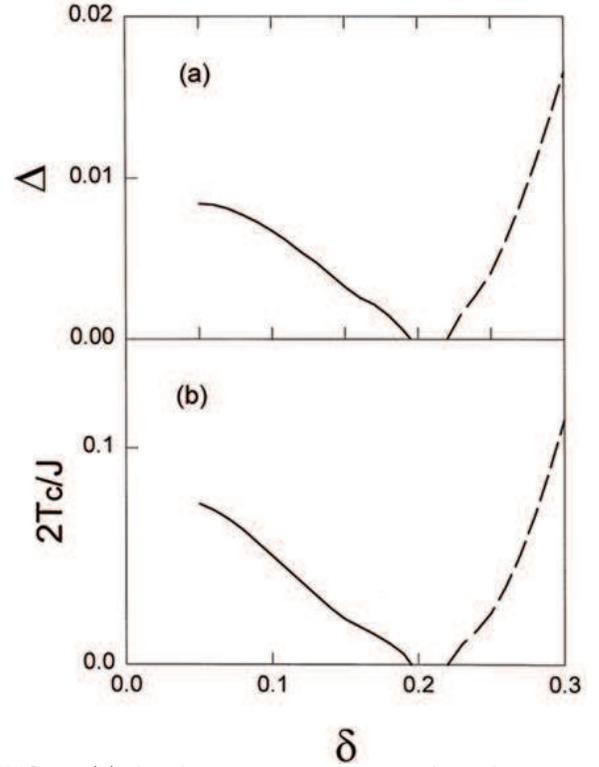}}
\caption{(a) the electron gap parameter $\Delta$ at
the zero temperature and (b) electron pair transition temperature
$T_{c}$ as a function of doping $\delta$ for $t/J=2$ in the s-wave
symmetry (solid line) and d-wave symmetry (dashed line).}
\end{figure}

\begin{figure}[prb]
\epsfxsize=3.0in\centerline{\epsffile{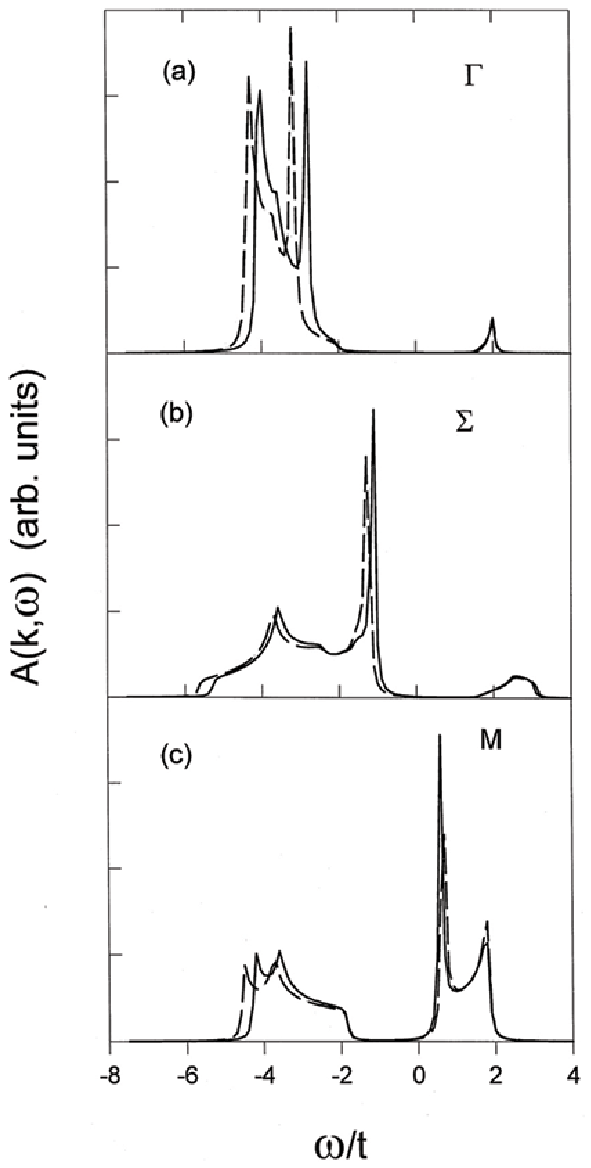}}
\caption{The electron spectral function $A({\bf k},\omega )$ of the
$t$-$J$ model in the cases of the existence of the local electron
pairs (solid line) and without the local electron pairs (dashed line)
at the doping $\delta=0.12$ for $t/J=2$.}
\end{figure}

\begin{figure}[prb]
\epsfxsize=3.0in\centerline{\epsffile{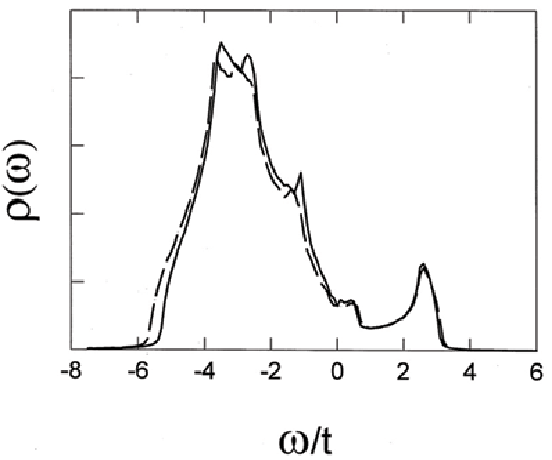}}
\caption{The electron spectral density of the $t$-$J$ model in the
cases of the existence of the local electron pairs (solid line) and
without the local electron pairs (dashed line) at the doping
$\delta=0.12$ for $t/J=2$.}
\end{figure}

For the further understanding the properties of this local electron
pairing state, we consider the electron spectrum function and electron
density of states. According to the previous discussions \cite{n14}, the
electron Green's function
$G({\bf k},t-t^{\prime})=\langle\langle C_{k\sigma}(t);C^{\dagger}_{k\sigma}
(t^{\prime})\rangle\rangle$, which is a convolution of the spinon Green's
function $D({\bf k},t-t^{\prime})$ and holon Green's function
$g({\bf k},t-t^{\prime})$, can be obtained in the present case as,
\begin{eqnarray}
G({\bf k},\omega )&=&{1\over N}\sum_{p}{\Lambda [(2\epsilon\chi_{z}
+\chi)\gamma_{p}-(\epsilon\chi+2\chi_{z})]\over 2\omega (p)}
\nonumber\\
&\times&\left\{ {1\over 2}\left (1+{\xi_{p-k}\over E_{p-k}}\right )
\left ({F_{1}({\bf k},{\bf p})\over \omega -\omega (p)+E_{p-k}}
\right . \right .\nonumber \\
&+& \left .\left .{F_{2}({\bf k},{\bf p})\over \omega +\omega (p)
+E_{p-k}}\right )\right . \nonumber \\
&+&\left .{1\over 2}\left (1-{\xi_{p-k}\over E_{p-k}}\right )
\left ({F_{1}({\bf k},{\bf p})\over\omega +\omega (p)-E_{p-k}}
\right .\right. \nonumber \\
&+& \left . \left .{F_{2}({\bf k},{\bf p})\over \omega -\omega
(p)-E_{p-k}}\right ) \right\}.
\end{eqnarray}
With the help of this electron Green's function, we can obtain the electron
spectrum function, $A({\bf k},\omega)=-2{\rm Im}G({\bf k},\omega)$ and
the electron density of states $\rho (\omega )={1\over N} \sum_{k}A({\bf k},
\omega)$. The results of the electron spectral function and
electron density of states at the doping $\delta =0.12$ for $t/J=2$
are shown in Fig. 3 (solid line) and Fig. 4 (solid line), respectively.
For comparison, the results of the electron spectral function and
electron density of states in the case of the electron gap parameter
$\Delta=0$ at the doping $\delta =0.12$ for $t/J=2$ are also shown in
Fig. 3 (dashed line) and Fig. 4 (dashed line), respectively. We
\cite{n14} have shown that the electron spectral function and electron
density of states in the case without these local electron pairs obtained
from the mean-field fermion-spin theory are qualitative consistent with
the numerical simulations \cite{n20} and experiments \cite{n21}. In the
present case, although the electron spectra function and electron density
of states are shifted slightly to low energy regime for $\omega<0$, but
the global feature of the electron spectral function and electron density
of states is almost same with these in the case of the electron gap
parameter $\Delta=0$, which shows that the normal state properties of the
copper oxide materials are dominated by holons moving in the background
of the spinon pair correlation. In fact, it has been shown from the
experiments \cite{n22} and theoretical discussions \cite{n15,n16}
that many problems in the normal state of the copper oxide materials
rely on the spinon pairing and does not require the existence of the
local pairs of holons.

In summary, we have discussed the physical properties of the electron
pairing state in the copper oxide materials within the fermion-spin
theory. According to the common form of the electron Cooper pair, we
show that there is the coexistence of the electron Cooper pair and
magnetic short-range correlation, and hence the AF short-range
correlation can persist into the superconducting state, which is
consistent with the experiments \cite{n6}. Within the mean-field level,
our results indicate that the electron pairing state originating from
the pure magnetic interaction in the 2D $t$-$J$ model is the local
state, and then does not reveal the true superconducting ground-state.

Finally, we note that many serious numerical studies carried out by
several groups \cite{n7,n192} showed that at the temperatures and lattice
sizes currently accessible to Monte Carlo simulations the pure 2D $t$-$J$
model does not superconduct at large $t/J$ and small hole density. Our
present result is consistent with these numerical results. Since the
boundary between the local electron s-wave pairing state and d-wave
pairing state is changed with the parameter $t/J$, we have also found
that there is a coexistence of the local electron s-wave pairing state
and d-wave pairing state in the underdoped and optimal doped regimes
for small $t/J$, and the physical properties of this pairing state is
dominated by the local electron d-wave pairing state, which is in
quantitative agreement with the numerical result \cite{n29}. In fact,
the d-wave gap function
$\Delta^{(d)}(k)\propto {\rm k}^{2}_{x}-{\rm k}^{2}_{y}$ belongs to
the same representation $\Gamma_{1}$ of the orthorhombic crystal group as
does s-wave gap function
$\Delta^{(s)}(k)\propto {\rm k}^{2}_{x}+{\rm k}^{2}_{y}$, the two perhaps
can mix at will, there are some evidences from the experiments to support
this symmetrical picture \cite{n30}. Although the normal-state is
two-dimensional and coherent transport in the c-axis is blocked due to
the absent of the coherent c-axis electron motion in the copper oxide
materials \cite{n26}, it is possible that the electron Cooper pairs
induced by the magnetic interaction within a given layer can tunnel
between interlayers freely by the Josephson mechanism, and then the
superconductivity may be motivated as the two- to three-dimensional
crossover \cite{n27}.

\acknowledgments

This work was supported by the National Natural Science
Foundation under Grant No. 19774014 and the State Education Department
of China through the Foundation of Doctoral Training. We thank Prof. H.
Q. Lin and Prof. G. S. Tian for bringing Ref. \cite{n192} to our attention.

\end{document}